# On the compression of the fullerene shell by an extra positive charge at its center

M.Ya. Amusia[1, 2], A.S. Baltenkov[3] and L.V. Chernysheva[2]

[1] Racah Institute of Physics, the Hebrew University, Jerusalem, 91904 Israel
[2] Ioffe Physical-Technical Institute, St. Petersburg, 194021 Russia
[3] Arifov Institute of Ion-Plasma and Laser Technologies,
Tashkent, 100125 Uzbekistan

**Abstract**

In this Letter, we investigate the variation of endohedral A@$C_N$ potential due to addition at the center of it a positive charge, for example, in the process of atom A photoionization. Using a reasonable model to describe the fullerenes shell, we managed to calculate the variation that is a consequence of the monopole polarization of $C_N$ shell.

We analyze model potentials with flat and non-flat bottoms and demonstrate that the phenomenological potentials that properly simulates the $C_{60}$ shell potential should belong to a family of potentials with a non-flat bottom. As concrete example, we use the Lorentz-bubble model potential. By varying the thickness of this potential, we describe the various degrees of the monopole polarization of the $C_{60}$ shell by positive electric charge in the center of the shell.

We calculated the photoionization cross-sections of He, Ar and Xe atoms located at the center of $C_{60}$ shell with and without taking into account accompanying this process monopole polarization of the fullerenes shell. Unexpectedly, we found that the monopole polarization do not affect the photoionization cross sections of these endohedral atoms.

**1.** The aim of this Letter is to find the extra potential induced in the electron shell of an endohedral A@$C_N$ under the action of an extra positive charge $z$ added at its center due to, for example, photoionization of atom A. We intend also to investigate the effect of this extra potential upon photoionization cross-section of an endohedral.

The idea that the electron interaction with almost fullerene $C_N$ can be described by a phenomenological potential $U(r)$ formed by smeared carbon atoms is a widely used approach (see, for example [1,2] and references therein) despite the fact that this approach is an essential simplification of the real molecular field. The atom smearing procedure determines the form of the potential function $U(r)$. If it is assumed [3] that the positive charge of the C nuclei together with the negative charge of the electrons is smeared in volume between two concentric spheres, we come to a function $U(r)$ in the form of potential well with a flat bottom. We can use another averaging. At first, let us smear the positive charge over a sphere with the radius $R$, taking into account that all the nuclei of carbon atoms are located at equal distances from the center of the fullerene cage. After that, we define the electron spatial distribution in the electric field of this sphere, and come to a potential well with a non-flat bottom [4].

Poisson equation determines the mutual spatial distribution of positive and negative electric charges creating the potential $U(r)$. The solutions of this equation are significantly different for these two types of the function $U(r)$. The functions with a flat bottom correspond to onion-like molecular structure with two spatially separated double-charges layers; while the functions $U(r)$ with non-flat bottom correspond to three-layer charges structure, in which the layer of positive charges is located between two negative ones. Such an arrangement of the positive and negative components of the fullerene shell leads to appearance of a minimum of the function $U(r)$ at $r=R$.

As a concrete example of $C_N$, let us consider almost spherical fullerene $C_{60}$. Its shell under the action of an external electric field, for example, a positive charge $z$[1] located far from the center of the cage, acquires an induced electric dipole moment and the $C_{60}$ shell becomes dipolar polarized. The $z$ charge in the center of $C_{60}$ compresses the electron cloud toward the center of the shell, retaining its spherical shape, thus leading to its monopole polarization. We encounter such a situation, for example, in photoionization of the endohedrals with atom A located at the center of the fullerenes cage.

The inner volume of $C_{60}$ shell is large enough to accommodate individual atoms A or even small molecules. The Van der Waals forces acting between the electrically neutral atom A and the $C_{60}$ shell are too weak to distort the electronic structure of both A and the shell itself, and therefore these structures one can consider as independent of each other. The situation changes because of, for example, photoionization of encapsulated atom [5], when a positive charge arises in the shell center. In this case, the positive electric field of atomic residue shifts the negative electron density of the $C_{60}$ shell relative to positive density of carbon ions. The shifting of the electron density in each elementary volume of the $C_{60}$ shell under the action of positive atomic residue $A^+$ results in creating an induced electric dipole moment of this volume, the axes of all elementary dipole moments being directed to the center of the $C_{60}$ sphere. As a result, the electric component of the shell, as a whole, shifts toward the sphere center, leading to monopole polarization of the shell and change of the shape of $C_{60}$ static potential $U(r)$.

Paper [5] analyzes the effect of monopole polarization in the process of $A@C_{60}$ photoionization. The authors of [5] called it "interior static polarization of the $C_{60}$ shell". They write: "The quintessence of the effect is that the ion reminder $A^+$, ones the photoionization taken place and the photoelectron is produced, could polarize the $C_{60}$ cage…This causes that the fullerene shell potential $U(r)$ to be different versus the situation without consideration of static polarization". To take into account this effect, the authors of [5] have introduced a modified version of rectangular potential well. They conclude that the interior static polarization of the $C_{60}$ shell "may not be ignored in the photoionization of endohedral atoms near threshold".

The present Letter is devoted to the consideration of $C_{60}$ monopole polarization within the framework of other type of model potentials with non-flat bottom, so-called the Lorentz-bubble potential [6, 7]. First, we examine with the Poisson equation the spatial distribution of charges that create a potential well with a Lorentz profile. Then, by varying the thicknesses of the left and right wings of the potential well, we establish the relationship of this variation with the displacement of part of the collectivized electrons relative to the positive frame of the fullerene shell. Further, we will estimate the relative number of electrons pulled into the center of the cage by a positive electric charge $z$ in its center. We will use the asymmetric Lorentz potential well to calculate the photoionization cross sections of He, Ar and Xe endohedral atoms.

**2.** The model potential $U(r)$ for $C_{60}$ shell should satisfy two general requirements. At first, it should be attractive and support at least one electronic s-state with the binding energy $E_s = -2.65 eV$, thus reproducing the electron affinity measured in UV photoelectron spectroscopy of the $C_{60}$ negative ion [8]. At second, the potential $U(r)$ should be located near the experimental fullerene radius $R$ in a rather thin spherical shell with thickness $\Delta$ of about few atomic units. In [6, 7] $U(r)$ has been calculated within the framework of the self-consistent spherical jellium model for collectivized electrons. In this approach, the electrostatic potential of the fullerene shell as a whole is a sum of the positive potential of carbon atom nuclei smeared over a sphere with radius $R$, and a negative potential created by the electron clouds. The parameters of this Lorentz-shape bubble potential is

$$U(r) = -U_{max} \frac{d^2}{(r-R)^2 + d^2}. \quad (1)$$

---
[1] We employ the atomic system of units (at. un.) in this paper.

In [7], the chosen depth $U_{max}$ and thickness $\Delta = 2d$ (at the middle of the maximal depth) permitted to locate in (1) an electronic level that corresponds to the experimental electron affinity of the $C_{60}$ molecule.

The relation $U(r) = -\varphi(r)$ connects Eq. (1) with the potential of electric field $\varphi(r)$ by which $C_{60}$ acts upon an electron. Here we take into account that the electron charge is equal to (-1). The Poisson equation [9] defines the electrostatic field potential $\varphi(r)$

$$\Delta\varphi = -4\pi\rho. \quad (2)$$

Here $\rho(r)$ is the density of the electric charges forming the potential well (1). In spherical coordinates with the center at the center of the $C_{60}$ cage the following equation defines the radial dependence of charge density:

$$\frac{1}{r}\frac{d^2}{dr^2}[rU(r)] = 4\pi\rho(r). \quad (3)$$

Applying (3) to potential (1), we obtain the spatial electric charge distribution $Q(r) = 4\pi\rho r^2$ that produces the Lorentz-shape bubble potential (1). Figure 1 (upper panel) presents the function $Q(r)$ versus radius $r$. The charge density in this figure is a three-layer sandwich. The middle layer represents the positively charges $C^{4+}$ ions, while the inner and outer layers represent the negatively charged clouds of collectivized electrons. The total charge of the $C_{60}$ shell is equal to zero because the potential (1) is a short-range potential. Integration of the negative parts of curve $Q(r)$ shows that about 46% of negative charge is located in the inner electronic cloud. The outer electronic cloud contains the rest negative charge of the $C_{60}$ shell.

Let us show that by varying the parameter $\Delta$ of the left and right wings of potential well (1) we can describe the transition of electrons from the outer electron cloud to the inner one through the rigid positive frame of the $C_{60}$ fullerene and analyze the various degrees of the monopole polarization of its shell. For this, we replace the constant thickness $\Delta$ in (1) by the following expression

$$\Delta(r) = \Delta_L + (\Delta_R - \Delta_L)\Theta(R-r). \quad (4)$$

Here $\Delta_L$ and $\Delta_R$ are parameters for the left and right wings of potential well (1), respectively; $\Theta(z)$ is the Heaviside step-function

$$\Theta(z) = [1+\exp(z/\eta)]^{-1}, \qquad (5)$$

in which the diffuseness parameter $\eta$ is a fixed positive number as small as we please, and which can therefore ultimately be replaced by zero. The lower panel of figure 1 gives electron densities of the left and right clouds versus radius $r$. The table presents the results of calculations of negative charges of these electron clouds. Note that the selection of the parameters is such that the affinity energy in the asymmetric Lorentz potential well remains always close to $E_s = -2.65 eV$.

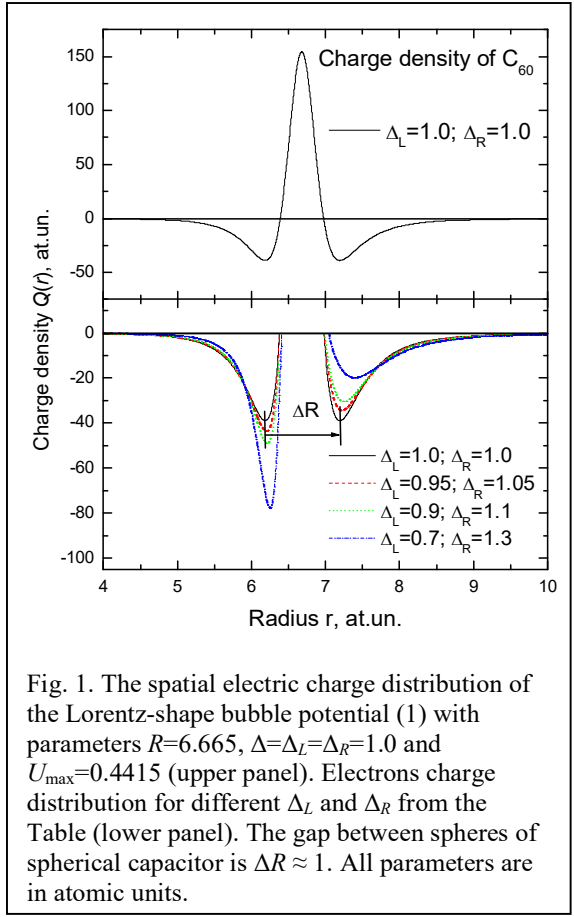

Fig. 1. The spatial electric charge distribution of the Lorentz-shape bubble potential (1) with parameters $R=6.665$, $\Delta=\Delta_L=\Delta_R=1.0$ and $U_{max}=0.4415$ (upper panel). Electrons charge distribution for different $\Delta_L$ and $\Delta_R$ from the Table (lower panel). The gap between spheres of spherical capacitor is $\Delta R \approx 1$. All parameters are in atomic units.

Let us estimate, in a crude approximation, the fraction of electrons transferred from the outer cloud to the inner one under the influence of a positive electric charge $z$ placed at the center of the cage. Suppose that the electrons of both clouds are located on concentric spheres with radii determined by the minima of the electron densities of the outer and inner clouds. According to figure 1, the gap between these spheres $\Delta R \approx 1$. We will consider the entire system of positive and negative charges as a spherical condensator, on the plates of which negative charges are placed in the proportions of 46% and 54%. We assume that such a positive charge between the spheres charge distributions stabilizes them. The electric field inside this condensator is equal to zero; otherwise, the electrons would have passed from one plate to another. After introducing the charge $z$ into the cage, some charge $q$ will move from the outer sphere to the inner one to compensate the electric field $z/R^2$. Taking into account that the electric field between the plates of a spherical capacitor is indistinguishable from the field of a point charge, one has $z/R^2 \approx q/\Delta R^2$. So, we obtain for the charge $q$ the following estimation $q \approx z(\Delta R/R)^2$. Thus, a unit charge $z=1$ causes the transfer of about 2% of electrons from the outer cloud to the inner one. This charge displacement corresponds to a set of the parameters $\Delta_L$ and $\Delta_R$ on the second row of the Table.

Table. Charges of the inner (left) and outer (right) electron clouds

| Parameters $\Delta_L$ and $\Delta_R$ | Charge of left wing, in % | Charge of right wing, in % |
|---|---|---|
| $\Delta_L=1.0$; $\Delta_R=1.0$ | 45.7 | 54.3 |
| $\Delta_L=0.95$; $\Delta_R=1.05$ | 48.2 | 51.8 |
| $\Delta_L=0.9$; $\Delta_R=1.1$ | 50.8 | 49.2 |
| $\Delta_L=0.7$; $\Delta_R=1.3$ | 61.2 | 38.8 |

Let us apply symmetric and asymmetric Lorentz potential wells to calculate the photoionization cross-sections of some endohedral atoms. The specific feature of an endohedral atom photoionization is oscillations in the photoionization cross-sections, commonly known as confinement resonances. Comparing confinement resonances for two types of the Lorentz potential wells, we shall evaluate the role of monopole polarization of the $C_{60}$ shell in this process. These potentials we have to add to the atomic Hartree-Fock Hamiltonian $\hat{H}_0$, thereby obtaining the equation for the electrons of endohedral atom located at the center of $C_{60}$ shell. Solutions of the following equation

$$[\hat{H}_0 + U(r)]\psi(\mathbf{r}) = E\psi(\mathbf{r}) \qquad (6)$$

give the initial- and final-state electron wave functions of encapsulated atom. They are used in calculations of photoionization cross-sections. The results are presented in figures 2-4.

The curves in these figures correspond to the set of parameters $\Delta_L$ and $\Delta_R$ from the fourth row of the Table. The sets of parameters on the second and third rows give the curves indistinguishable from the curves corresponding to the symmetric potential well. Thus, the monopole polarization of the $C_{60}$ shell begins to affect the photoionization of the endohedral atom with a shift of approximately 15% of collectivized electrons. With this shift, the discrepancies of the curves for the symmetric and asymmetric potentials are of the order of the line thickness. Coincidence of curves demonstrates the absence of the monopole polarization effect on the photoionization of endohedral atom[*]. So, given in [5] statement that static monopole polarization of the $C_{60}$ shell "may not be ignored in the photoionization of endohedral atoms near threshold" is emphatically incorrect.

**3.** Paper [5] considers, as far as we know, the monopole polarization of the $C_{60}$ shell for the first time. To describe this effect, we employ a modified version of the spherical rectangular potential well. It has the following form

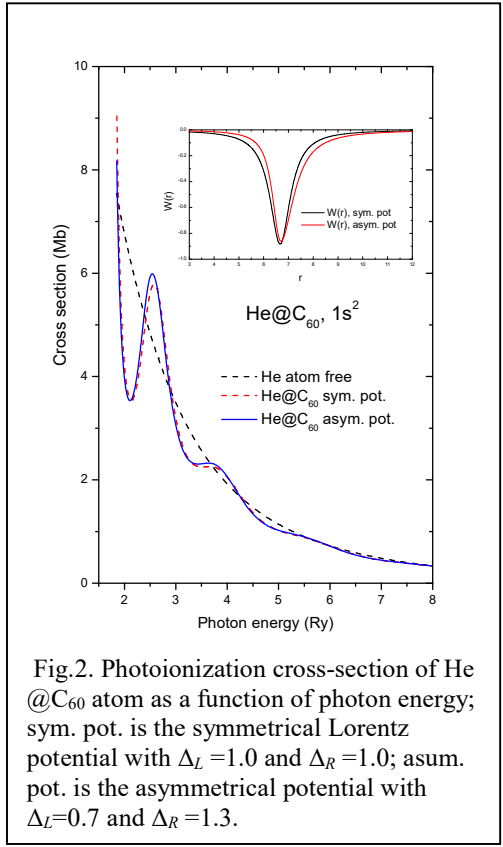

Fig.2. Photoionization cross-section of He @$C_{60}$ atom as a function of photon energy; sym. pot. is the symmetrical Lorentz potential with $\Delta_L$ =1.0 and $\Delta_R$ =1.0; asum. pot. is the asymmetrical potential with $\Delta_L$ =0.7 and $\Delta_R$ =1.3.

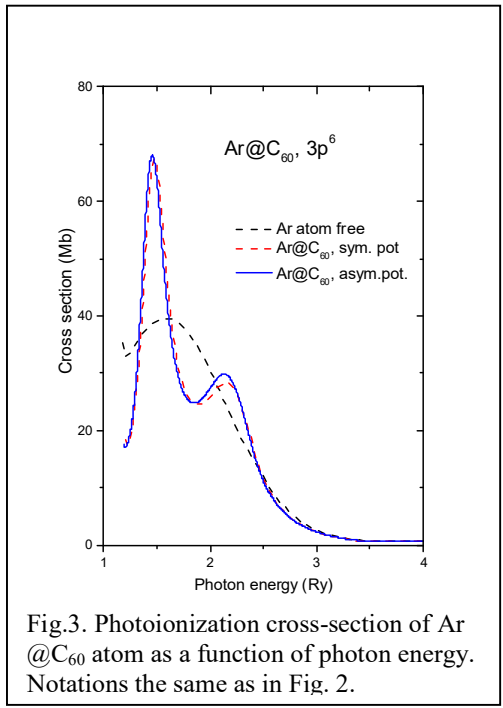

Fig.3. Photoionization cross-section of Ar @$C_{60}$ atom as a function of photon energy. Notations the same as in Fig. 2.

---

[*] Note that to the same conclusion we came on example of another potential $U(r)$ with non-flat bottom [10]

$$U^*(r) = \begin{cases} \dfrac{\alpha}{r_0} - \dfrac{\alpha}{r_0 + \Delta r}, & \text{if } r \leq r_0; \\ -U_0 + \dfrac{\alpha}{r} - \dfrac{\alpha}{r_0 + \Delta r}, & \text{if } r_0 \leq r \leq r_0 + \Delta r; \\ 0, & \text{otherwise.} \end{cases} \quad (7)$$

Here $r_0$ denotes the inner fullerenes radius, $\Delta r$ is the thickness, and $U_0$ is the depth of the $C_{60}$ potential well; parameter $\alpha$ is equal either to zero, $\alpha=0$, or to 1, if the monopole static polarization is entirely ignored or complete included, respectively.

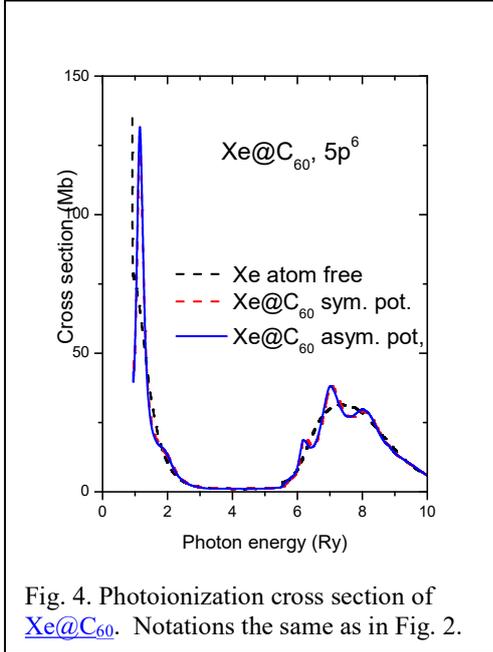

Fig. 4. Photoionization cross section of Xe@C$_{60}$. Notations the same as in Fig. 2.

It is interesting to trace the emergence and development of the concept of rectangular potential itself. In one of the first papers [2] we read: "We introduce a shell of positive rigid background charge, jellium, which is symmetrically placed with respect to the radius $R$ of the $C_{60}$ molecule"; i.e., the positive charge of the shell is uniformly filling the spherical layer between two concentric spheres. If the authors [2] suggested that the electrons of the shell are located in it, like "raisins in a bun", then they would have the structure resembling the J.J. Thomson's "plum pudding model" of an atom. However, the authors of [2] went another way and defined the electron density in the shell and its effective potential by solving the Poisson equation with rigid positive spherical layer. As a result, they obtained the square potential well that now is a widely used model for the $C_{60}$ shell without any mention about the assumptions made.

Let us apply the Poisson equation (3) to potential function (7) to understand, what charge distribution corresponds to this potential well. Using the Heaviside step function (5), we rewrite the function (7) in the following form

$$U^*(r) = \left(\dfrac{\alpha}{r_0} - \dfrac{\alpha}{r + \Delta r}\right)\Theta(r - r_0) - \left(U_0 - \dfrac{\alpha}{r} + \dfrac{\alpha}{r_0 + \Delta r}\right)\Theta(r_0 - r)\Theta(r - r_0 - \Delta r). \quad (8)$$

The reasons why we replace the stepwise function (7) by diffuse potentials (8) will become clear later. Let us begin from the case $\alpha = 0$ that corresponds to the spatial distribution of the electric charge densities when the static polarization of the shell is ignored (the details of calculations are in [11]). Figure 5 presents the charge distribution function $Q(r) = 4\pi\rho r^2$. According to this figure, two concentric spheres with radii $r = r_0$ and $r_0+\Delta r$ bearing the double electric layers create the radial non-modified square well potential. The thicknesses of layers (for parameter $\eta = 0.05$) are about 0.05 at. un. Both spheres are electrically neutral. On a surface of the inner sphere, about 36% of positive and negative charges are located. The surface of the outer sphere localizes the rest of the $C_{60}$ shell charges. For potential well without diffuseness ($\eta = 0$), the charge layers have zero thickness. The function $Q(r)$ is equal to zero everywhere except the points $r = r_0$ and $r_0+\Delta r$, in the infinitesimal vicinity of which the charge densities are infinitely negative and positive.

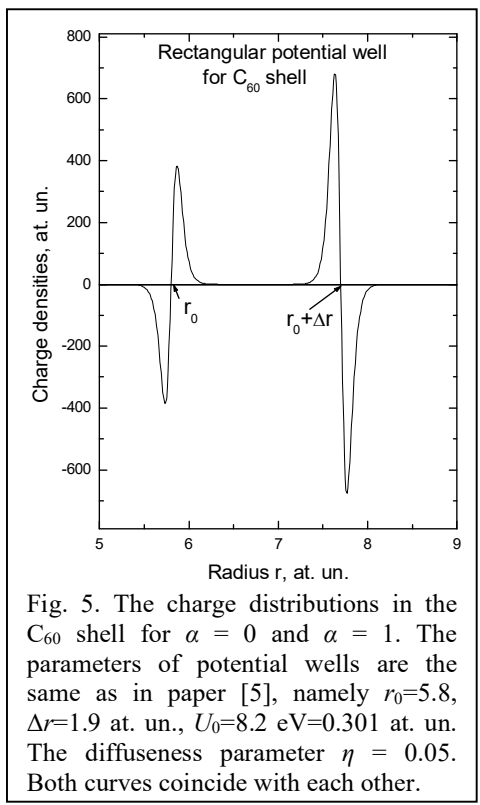

Fig. 5. The charge distributions in the $C_{60}$ shell for $\alpha = 0$ and $\alpha = 1$. The parameters of potential wells are the same as in paper [5], namely $r_0$=5.8, $\Delta r$=1.9 at. un., $U_0$=8.2 eV=0.301 at. un. The diffuseness parameter $\eta = 0.05$. Both curves coincide with each other.

Repeating the same procedure with potential function (8) for $\alpha = 1$ we come to the same charge distribution as in the case $\alpha = 0$. The reason for such an unexpected, at the first glance, result is as follows. Let us apply the Laplacian $\Delta$ from the Poisson equation (3) to additional terms in Eq.(7). For the first line we have

$$\Delta\left(\frac{\alpha}{r_0} - \frac{\alpha}{r_0 + \Delta r}\right) \equiv 0. \qquad (9)$$

For the second line, since the Coulomb potential $\alpha/r$ is the Green function of the Poisson equation [9], it is

$$\Delta\left(\frac{\alpha}{r} - \frac{\alpha}{r_0 + \Delta r}\right) = -4\pi\alpha\delta(\mathbf{r}), \qquad (10)$$

. Again, we have zero in the right side of (10) because $\mathbf{r}\neq 0$ in this line. Thus, the additional terms in potential function (7) do not lead to alterations in the mutual disposition of electric charges in the $C_{60}$ shell, as well as to static monopole polarization of the fullerene shell by the electric charge located in the center of the shell. Therefore, the question arises, what is the reason for the significant change in the photoionization cross section predicted in [5]. The answer is that the authors of [5] simply introduced new arbitrary parameter (in addition to old arbitrary parameters $r_0$ and $\Delta r$) into the ordinary rectangular potential, which have nothing to do with the monopole polarization of the $C_{60}$ shell.

**4.** We observe here an interesting and strange peculiarity: the monopole polarization of an endohedral shell that appear due to inner atom photoionization does not affect the cross section of this process. The result is obtained, however, in the frame a rather crude model that describes the fullerenes electron shell.

The problem with model description of the shape and parameters of the fullerene shell potential is similar, to some extent, to that in nuclear physics where the potential for the nucleon-nucleon interaction is unknown. To describe the nuclei the researchers use complicated shapes of the potentials, up to such exotic as "Elsasser wine bottle" or "Mexican hat" depending on a great number of parameters. In such a way, one could model all the magic numbers of nuclei (see e.g. [12]). We hope that more detailed experimental investigations of $C_{60}$ itself and endohedral systems A@$C_{60}$ will make the search for the model potential of $C_{60}$ more constructive than a simple increase in the number of arbitrary parameters in the old model and will discover a new avenue to modification of the $C_{60}$ shell models.

**Acknowledgments**
ASB is grateful for the support to the Uzbek Foundation Award OT-Ф2-46.

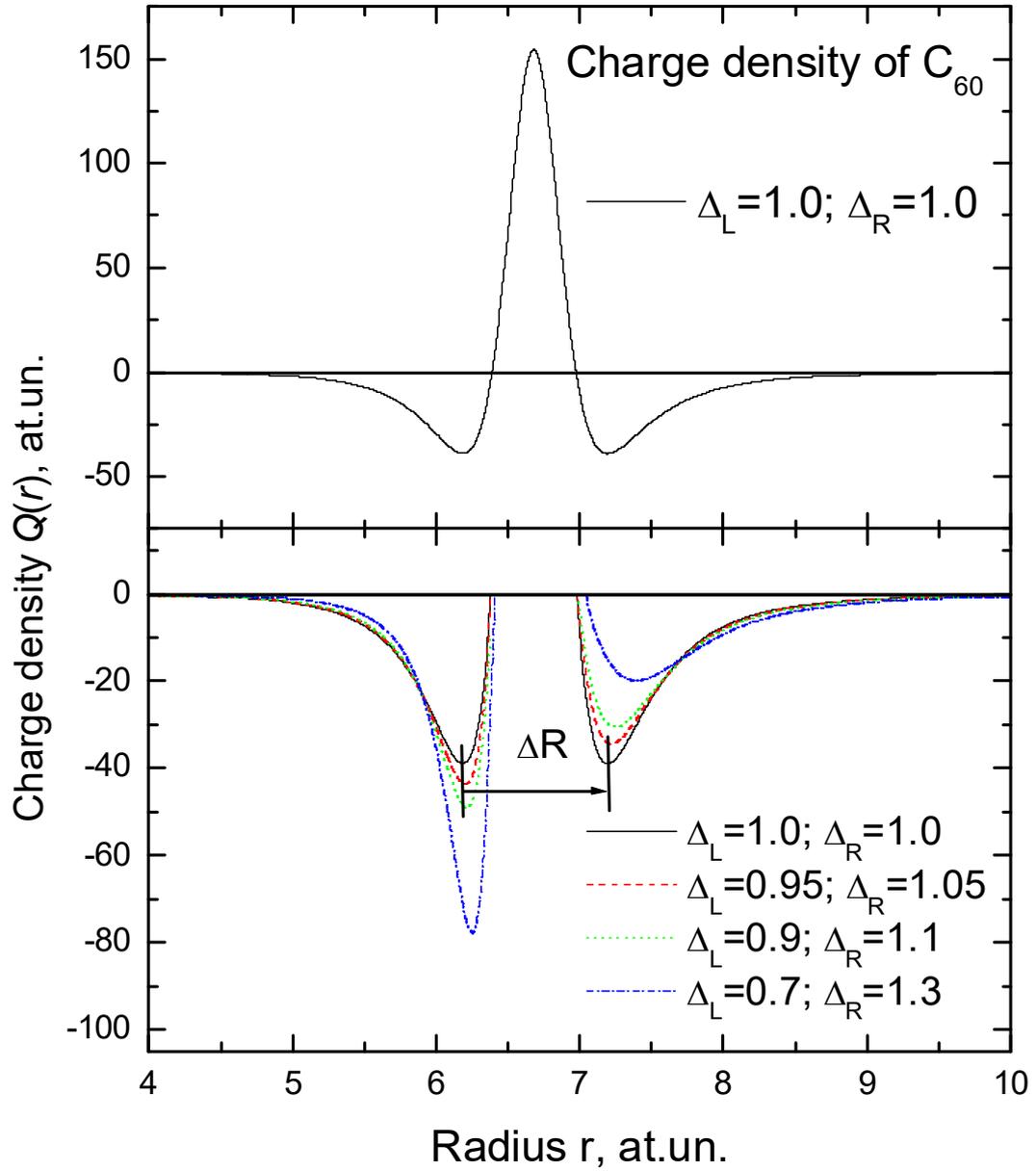

Fig. 1. The spatial electric charge distribution of the Lorentz-bubble potential (1) with parameters $R=6.665$, $\Delta=\Delta_L=\Delta_R=1.0$ and $U_{max}=0.4415$ (upper panel). Electrons charge distribution for different $\Delta_L$ and $\Delta_R$ from the Table (lower panel). The gap between sphere of spherical capacitor is $\Delta R \approx 1$. All parameters are in atomic units.

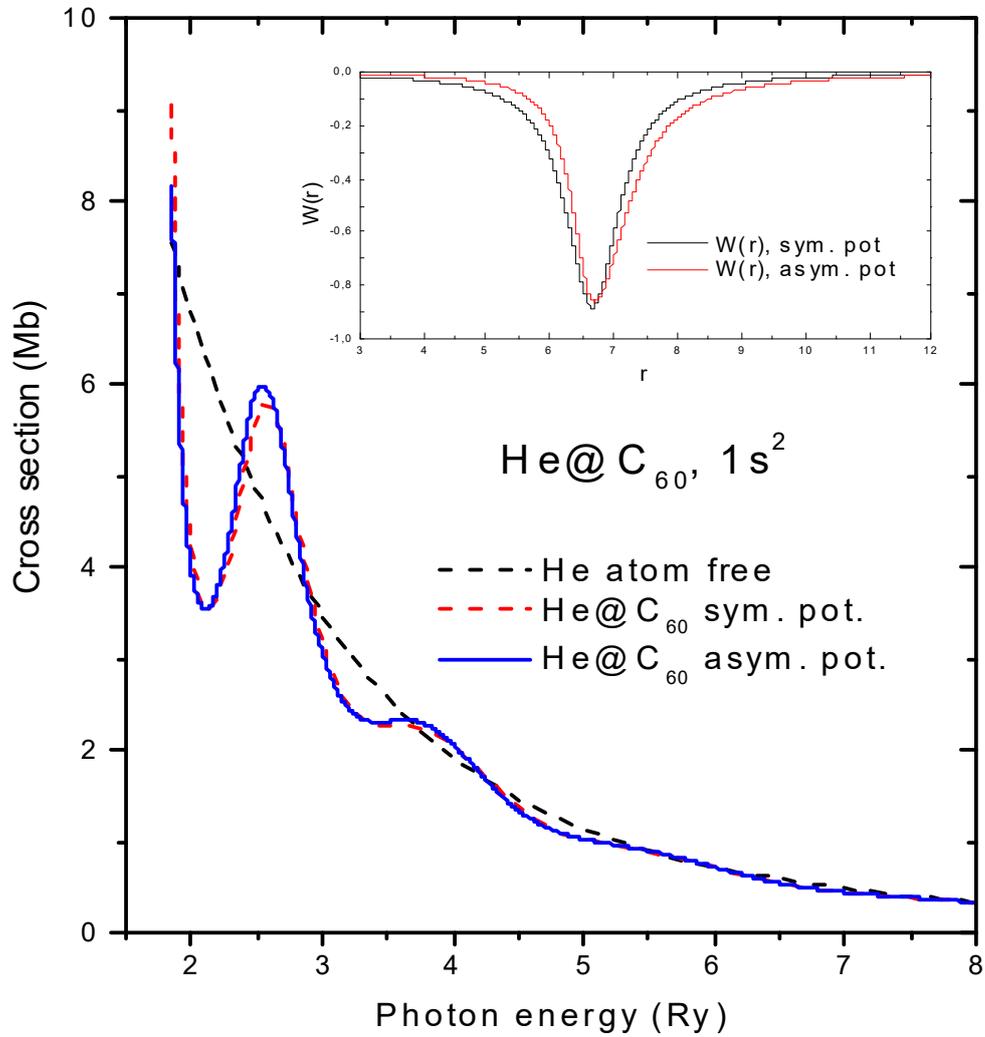

Fig. 2. Photoionization cross section of endohedral He atom as a function of photon energy; sym. pot. is the symmetrical Lorentz potential with $\Delta_L = 1.0$ and $\Delta_R = 1.0$; asum. pot. is the asymmetrical potential with $\Delta_L = 0.7$ and $\Delta_R = 1.3$.

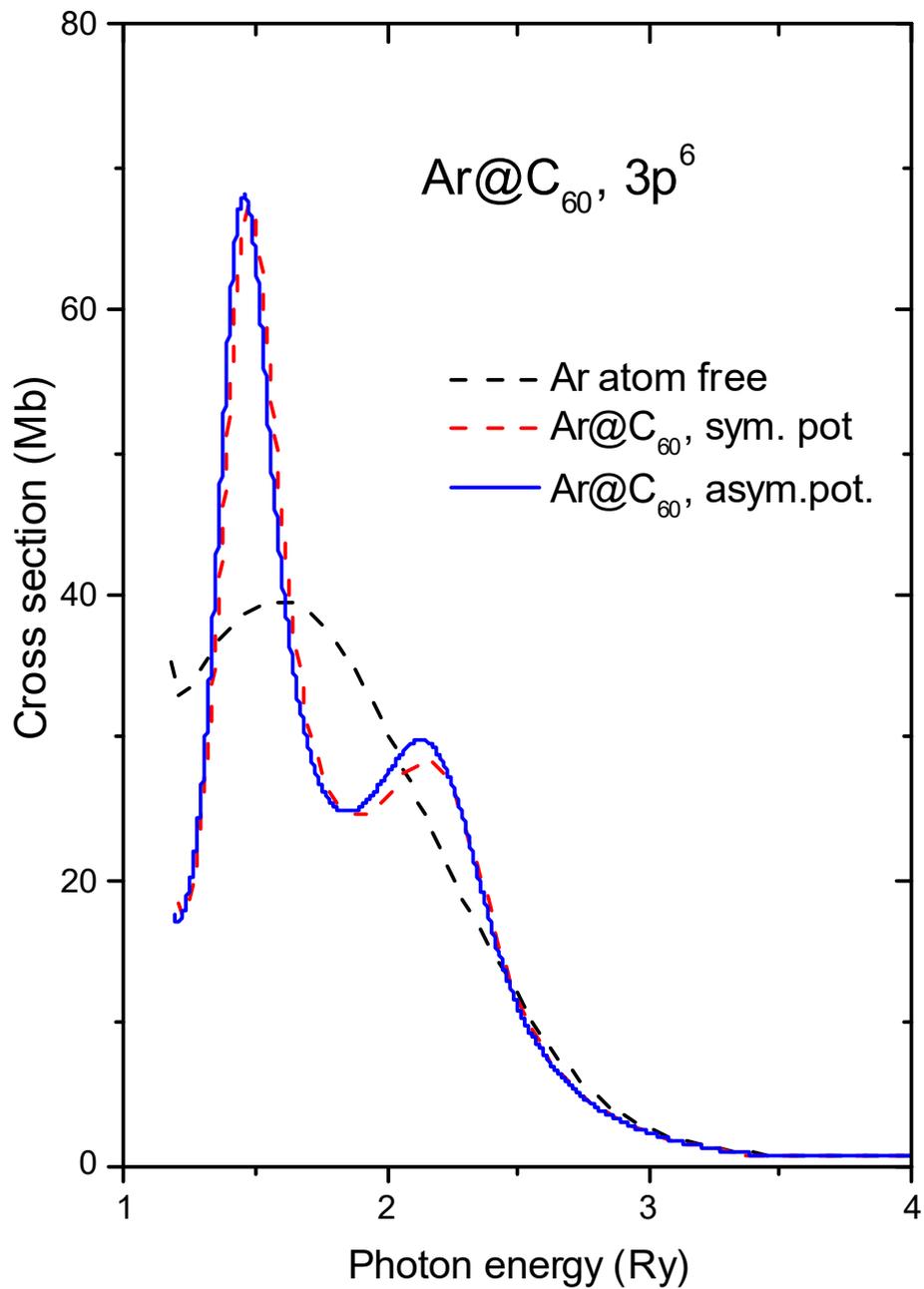

Fig.3. Photoionization cross section of endohedral Ar atom as a function of photon energy; sym. pot. is the symmetrical Lorentz potential with $\Delta_L = 1.0$ and $\Delta_R = 1.0$; asum. pot. is the asymmetrical potential with $\Delta_L = 0.7$ and $\Delta_R = 1.3$.

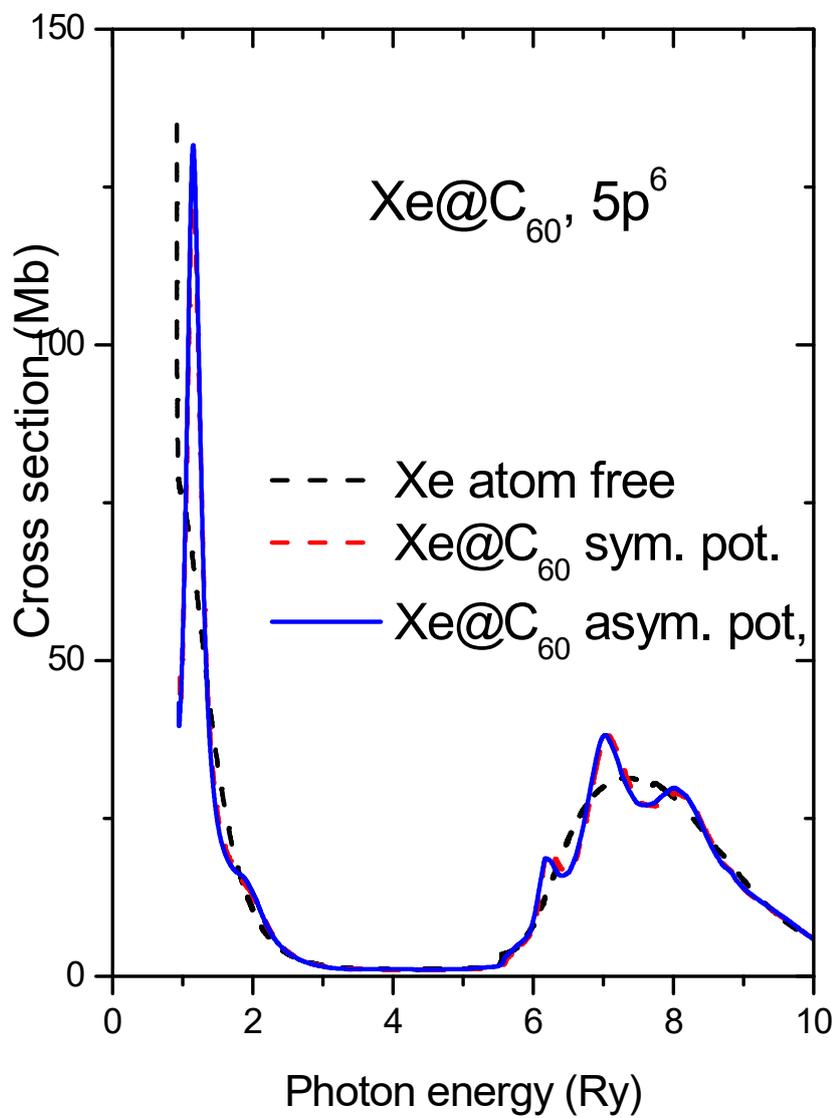

Fig. 4. Photoionization cross section of endohedral Xe atom as a function of photon energy; sym. pot. is the symmetrical Lorentz potential with $\Delta_L =1.0$ and $\Delta_R =1.0$; asym. pot. is the asymmetrical potential with $\Delta_L=0.7$ and $\Delta_R =1.3$.

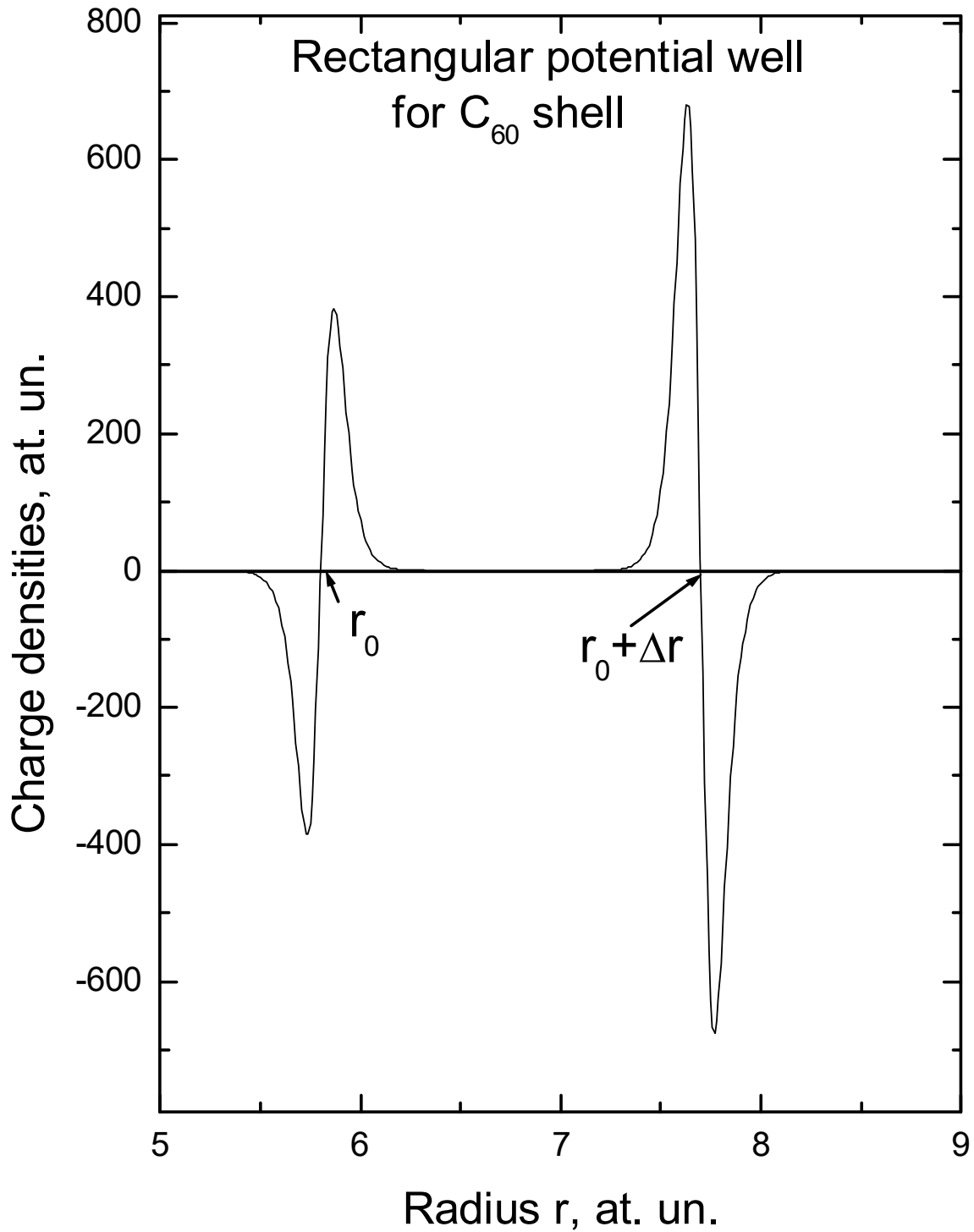

Fig. 5. The charge distributions in the C$_{60}$ shell for $\alpha = 0$ and $\alpha = 1$. The parameters of potential wells are the same as in paper [5], namely $r_0$=5.8, $\Delta r$=1.9 at. un., $U_0$=8.2 eV=0.301 at. un. The diffuseness parameter $\eta = 0.05$. The both curves coincide with each other.